\documentclass[pra,twocolumn,groupedaddress,showpacs,amsmath,floatfix]{revtex4-1}
\usepackage{graphicx}
\usepackage{dcolumn}
\usepackage{bm}
\usepackage{amsmath}
\usepackage{amssymb}
\usepackage{revsymb4-1}

\begin{document}

\title{Nonlinear properties and stabilities of polaritonic crystals beyond the low-excitation-density limit}

\author{E. S. Sedov}
\author{A. P. Alodjants}
\email[Electronic address: ]{alodjants@vlsu.ru}
\author{S. M. Arakelian}
\affiliation{Department of Physics and Applied Mathematics, Vladimir State University, Gorky Street 87, RU-600000 Vladimir, Russia}

\author{Y.Y. Lin}
\author{R.-K. Lee}
\affiliation{Institute of Photonics Technologies, National Tsing-Hua University, Hsinchu 300, Taiwan}


\begin{abstract} Coherent properties of a  two-dimensional spatially periodic structure, polaritonic crystal (PolC) formed by trapped two-level atoms in an optical cavity array interacting with a light field, are analyzed. By considering the wave function overlapping for both photonic and atomic states, a cubic-quintic complex nonlinear Schr\"odinger equation is derived for the dynamics of coupled atom-light states, wave function of low-branch polaritons, associated with PolC in the continuous limit. A variational approach predicts that a stable ground-state wave function of PolC exists but is accompanied by an oscillating width. For a negative scattering length, the wave function collapses in the presence of a small quintic nonlinearity appearing due to a three-body polariton interaction. By studying the nonequilibrium (dissipative) dynamics of polaritons with adiabatic approximation, we have shown that the collapse of PolC wave function can be  prevented even in the presence of small decaying of a number of polariton particles.
\end{abstract}
\maketitle

\section{\label{intro}INTRODUCTION}
Present remarkable achievements with ultracold trapped atomic gases evoke a great interest in investigating quantum phases for coupled matter-field states  \cite{Greiner02,SilverHohenadlerBhaseen}. By cooling atoms to about absolute zero, the state of matter known as Bose-Einstein condensate (BEC) is described as a macroscopic wave function that can extend over several micrometers \cite{BEC-book}. With optical lattices, artificial crystals made by interfering laser beams, one can observe many-body dynamics from a Mott-insulator (MI) phase to a superfluid (SF) phase in the gas of ultracold atoms with periodic potentials [1,4]. Recently, it has been demonstrated that the superradiant phase corresponds to the periodical self-organized phase of the atoms when a standing-wave laser-driven BEC is loaded into a high-finesse optical cavity \cite{Baumann2010p1301,Nagy2010p130401}.

Instead of a single cavity, state-of-the-art fabrication technology makes it possible to create periodical structures on the basis of coupled microcavity chains, where few-level atoms are placed inside \cite{LeeYariv,VahalaNature003}. With an array of optical cavities, atoms strongly interacting with photon modes can provide a platform to study quantum phase transitions of light by including photon-atom on-site interactions and photon hopping effects between adjacent cavities~\cite{Greentree06,Hartmann06,Angelakis06,Lei08}. Based on atom-light interactions in cavity arrays, exotic quantum states of light have been predicted for a Heisenberg spin-1/2 Hamiltonian \cite{SilverHohenadlerBhaseen,Hartmann2007p160501}, a two-species Bose-Hubbard model \cite{Hartmann2008p033011}, arrays of coupled cavities \cite{Rossini2007p186401,Angelakis2007p031805}, and dual-species optical-lattice cavities~\cite{Lei2010}. These studies allow us to analyze critical quantum phenomena in conventional condensed matter systems by manipulating the interaction between photons and atoms. In this respect, polaritons, bosonic quasiparticles,  representing a linear superposition of photons in the external electromagnetic field and excitations in a two-level system act as natural objects for the study of photon-atom interactions. Such coupled matter-field states have attracted lots of attentions in quantum physics for simultaneously possessing coherent matterlike and photonlike wave functions (cf. \cite{DengHaugYamamoto}).

At present, evidence of coherent macroscopic properties of polaritons has been found both in the solid state and the atomic physics domain --- see~\cite{KasprzakRichardKundermann,BaliliHartwellSnoke,UtsunomiyaTianRoumpos, ChenLuWu2011,PaternostroAgarwalKim,FleischauerLukin,LKarpaMWeitz,ChestnovAlodjantsArakelian, AlodjantsChestnovArakelianPRA2011,AlodjantsBarinovArakelian,FleischhauerOtterbachUnanyan, AlodjantsArakelianLeksinLaserPhysics}. First, we are speaking about promising experiments aimed at observing BEC phenomenon and SF properties of low-branch (LB) exciton polaritons in semiconductor quantum well structures embedded in Bragg microcavities \cite{KasprzakRichardKundermann,BaliliHartwellSnoke,UtsunomiyaTianRoumpos,ChenLuWu2011,PaternostroAgarwalKim}. In particular, macroscopic occupation of LB polaritons in Cd/Te/CdMgTe microcavities at a temperature 5K has been demonstrated in \cite{KasprzakRichardKundermann}. The first-order coherence and spontaneous linear polarization of light emission have been shown for polaritons trapped in harmonic potential \cite{BaliliHartwellSnoke}. Such behavior of polaritons opens the door to the investigation of many-body physics, solitons and pattern formation due to nonlocal nonlinear effects in matter-field interaction (cf.~\cite{UtsunomiyaTianRoumpos,PaternostroAgarwalKim}). Alternatively, in atomic optics the macroscopic coherent properties with atomic polaritons are observed in various problems of atom-field interaction where long-lived coherence of the quantized optical field strongly coupled with two (or multi)-level atoms can be achieved (cf. \cite{FleischauerLukin,FleischhauerOtterbachUnanyan,LKarpaMWeitz,AlodjantsChestnovArakelianPRA2011, AlodjantsBarinovArakelian,ChestnovAlodjantsArakelian,AlodjantsArakelianLeksinLaserPhysics}).

From a practical point of view such systems with coupled matter-field states (dark-state polaritons) represent an indispensable ingredient for designing temporary quantum memory and quantum-information processing devices. Obviously, in the real world such polaritonic devices should operate with a large number of qubits, which implies a large enough number of  cavities as well. We emphasize two important circumstances that must be taken into account if we want to implement spatially periodic structures for both phase transition problems and quantum computing purposes.
First, it is important to achieve a thermodynamic limit considering a macroscopically large number of cavities and small decay (decoherence) rates (cf.~\cite{AlodjantsBarinovArakelian}). 
Second, nonlinear effects arising due to polariton-polariton scattering should be taken into account in a general case.

Combining coupled quantum electrodynamic cavity arrays and ultracold atoms, we analyze trapped two-level atoms interacting with the photon fields in a two-dimensional (2D) cavity array at a zero temperature limit. Taking into consideration both photonic and atomic wave function overlapping between adjacent cavities, we introduce a polaritonic crystal (PolC) formed by the superposition of photonic cavity modes and atomic excitations in spatially periodic structures \cite{AlodjantsBarinovArakelian}. Based on the Holstein-Primakoff transformation \cite{HP} but being beyond the low-excitation-density approximation, we derive a cubic-quintic complex nonlinear Schr\"odinger equation (CNLSE) for LB polaritons in the continuum limit. We use the Gaussian variational approach to analyze the stability of a PolC structure for different atom numbers, two- and three-body polariton-polariton interaction strengths. The applicability of the variational method that is widely used for describing atomic BEC \cite{BEC-book,PerezGarciaMichinel,AbdullaevGammalTomio} (or optical beams and solitary waves \cite{QuirogaTeixeiroMichinel,KivsharAgrawalOpticalSolitons}) is justified if the shape of the actual solution of the NLSE is closer to the ansatz function.

The paper is arranged as follows. In Sec.~\ref{themodelofpolaritonic} we describe a model to realize 2D PolCs, that occur due to the atom-field interaction in a cavity array. Some aspects of many-body physics for PolCs in the momentum representation including their three-body interactions are established in Sec.~\ref{polaritonpropertiesinmomentum}. In Sec.~\ref{nonlineardynamicsofpolaritons}, we study the dynamics of PolCs in the continuum limit, where the corresponding stabilities of a ground-state wave function at equilibrium are shown by the variational approach. Nonequilibrium effects of a weak polariton number decaying in the PolC structure are examined as well. In Sec.~\ref{conclus} we summarize our results.
 
\section{\label{themodelofpolaritonic}THE MODEL OF PolC}

\begin{figure}
\includegraphics[width = 8.3cm]{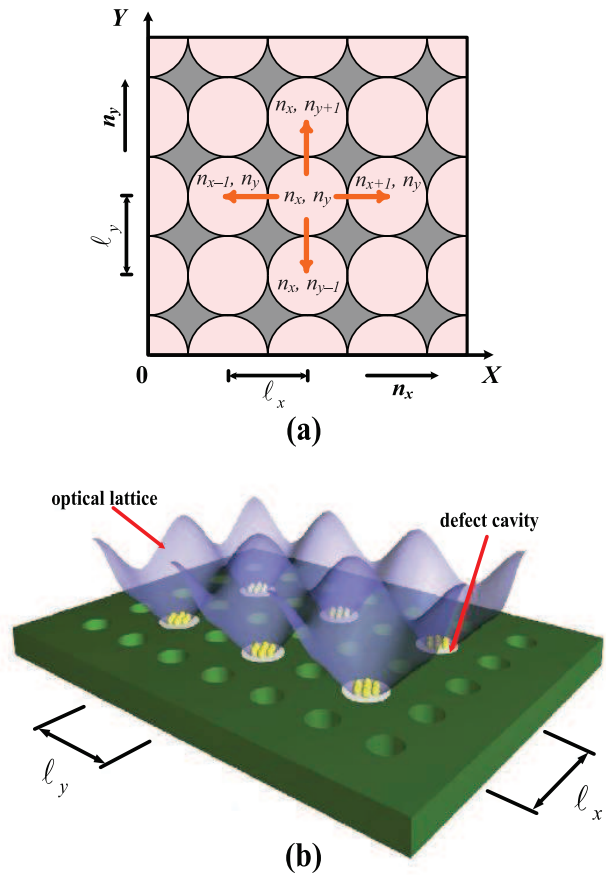}
\caption{\label{PicLat} (Color online) (a) Schematic picture of the proposed PolC in a 2D structure. Each cavity lattice has a nearest-neighbor interaction in the $XY$ plane. The integer numbers $n_{x} $ and $n_{y} $ enumerate lattice cells containing cavities; ${n_{x} = 1, 2,\dots,N_{x} }$ and ${n_{y} =1, 2,\dots,N_{y}}$, with the number of cavities in the $X$ and $Y$ directions as $N_{x} $ and $N_{y} $, respectively. $M = N_{x} \times N_{y}$ is the total number of cavities.
(b) Illustration of a possible PolC system formed by loading ultracold atoms into optical lattices within a photonic defect cavity array.}
\end{figure}

PolC structure can be created basically by means of atom-field interaction in a 2D cavity array, as illustrated in Fig.~\ref{PicLat}.
Here we consider the array of $M$ single-mode micro-cavities with the nearest-neighbor interactions in the $XY$ plane.
Each of the cavities represents an {atom-photon cluster} system, which contains a small but macroscopic number of ultracold two-level atoms with two internal states labeled as ${\left| a \right\rangle} $ and ${\left| b \right\rangle} $, respectively.
To produce such a system experimentally, one may trap ultracold two-level atoms in 2D optical lattices the minimum of which coincide with positions of defect cavities in a band-gap structure, as shown in the schematic picture in Fig.~\ref{PicLat}(b). We can represent the total Hamiltonian for the system in Fig.~\ref{PicLat} as
\begin{equation}
\label{HamTot}
\hat{H}=\hat{H}_{at}+\hat{H}_{ph}+\hat{H}_{int},
\end{equation}
where $\hat{H}_{at}$ is a Hamiltonian for weakly interacting two-level atoms, $\hat{H}_{ph}$ is responsible for the photonic field distribution, and $\hat{H}_{int}$ characterizes the atom-light interaction in each cavity. These Hamiltonians can be written in the second quantized form as

\begin{subequations}
\label{HAPI}
\begin{eqnarray}
&&\hat{H}_{at} =\sum_{
\begin{smallmatrix}i,j=a,b\\i\ne j
\end{smallmatrix}}
\int \hat{\Phi}_{j}^{\dag} \left(-\frac{\hbar^{2} \Delta}{2M_{at}} +V_{ext}^{(j)} \right. \nonumber\\
&& \phantom{\hat{H}_{at}}+\left. \frac{1}{2}U_{j} \hat{\Phi}_{j}^{\dag} \hat{\Phi}_{j}+\frac{1}{2} U_{ab} \hat{\Phi}_{i}^{\dag}\hat{\Phi}_{i} \right)\hat{\Phi}_{j} d^{3}\mathbf{r},\label{HAPIa}
\\
&&\hat{H}_{ph}=\int \hat{\Phi}_{ph}^{\dag} \left(-\frac{\hbar^{2} \Delta}{2M_{ph}} +V_{ph} \right) \hat{\Phi}_{ph} d^{3}\mathbf{r},\label{HAPIb}
\\
&&\hat{H}_{int}=\hbar \kappa \int \left(\hat{\Phi}_{ph}^{\dag} \hat{\Phi}_{a}^{\dag} \hat{\Phi}_{b} +\hat{\Phi}_{b}^{\dag} \hat{\Phi}_{a} \hat{\Phi}_{ph} \right) d^{3}\mathbf{r},\label{HAPIc}
\end{eqnarray}
\end{subequations}
where $M_{at}$ is a mass of free atoms and $M_{ph}$ is an effective mass of trapped cavity photons. The quantum field operators $\hat{\Phi}_{a,b}\left(\mathbf{r}\right)\,\text{and}\, \hat{\Phi}_{a,b}^{\dag}\left(\mathbf{r}\right)$ $\left(\hat{\Phi}_{ph}\left(\mathbf{r}\right)\,\text{and}\, \hat{\Phi}_{ph}^{\dag}\left(\mathbf{r}\right)\right)$ annihilate and create atoms (photons) at the position $\mathbf{r}$; while $V_{ext}^{(j)}$ ($j=a,b$) and $V_{ph}$ are the trapping potentials for the atoms and photons, respectively.
As an example, for the PolC structure illustrated in Fig.~\ref{PicLat}(b), the potential $V_{ext}^{(j)} $ for a magneto-optical trap can be chosen as $V_{ext}^{(j)} =V_{0} \left[\sin ^{2} \left(\frac{\pi x}{l_{x}} \right)+\sin^{2} \left(\frac{\pi y}{l_{y}} \right)\right]+\frac{1}{2} M_{at} \omega _{z} z^{2}$ with the optical lattice constants, $l_{x,y}$, and a characteristic frequency of harmonic trapping for atoms in $z$ direction, $\omega _{z}$. The interaction strength between two-level atoms and the quantized field is denoted by $\kappa$.

In Eq.~(\ref{HAPI}a) parameters ${U_a=\frac{4\pi \hbar^2 a_a^{(sc)}}{M_{at}}}$ and ${U_b=\frac{4\pi \hbar^2 a_b^{(sc)}}{M_{at}}}$ characterize atom-atom scattering processes at two internal levels, ${\left| a \right\rangle} $ and ${\left| b \right\rangle}$, respectively. Parameter $U_{ab}=\frac{4\pi \hbar^2 a_{ab}^{(sc)}}{M_{at}}$ is relevant to interactions between atoms in different internal states; $a_{a,b}^{(sc)}$ and $a_{ab}^{(sc)}$ are scattering lengths for corresponding elastic collisions of atoms (cf.~\cite{BEC-book}).

In general, one can expand the field operators $\hat{\Phi}_{a,b} (\mathbf{r})$ and $\hat{\Phi}_{ph} (\mathbf{r})$ by separable spatial wave functions as follows: 
\begin{subequations}
\label{OperFI}
\begin{eqnarray}
&&\hat{\Phi}_{a}(\mathbf{r})=\sum_{n_{x}, n_{y}}\hat{a}_{n_{x} n_{y}} \varphi_{n_{x} n_{y}}^{(a)}\left(\mathbf{r}\right),\label{OperFIa}\\
&&\hat{\Phi}_{b} (\mathbf{r})=\sum_{n_{x}, n_{y}}\hat{b}_{n_{x} n_{y}} \varphi_{n_{x} n_{y}}^{(b)} \left(\mathbf{r}\right),\label{OperFIb}\\
&&\hat{\Phi}_{ph} (\mathbf{r})=\sum_{n_{x}, n_{y}}\hat{\psi}_{n_{x} n_{y}} \xi_{n_{x} n_{y}} \left(\mathbf{r}\right),\label{OperFIc}
\end{eqnarray}
\end{subequations}
where $\varphi_{n_{x} n_{y}}^{(a,b)}$ and $\xi_{n_{x} n_{y}}$ are real Wannier functions representing spatial distributions of ultracold atoms and photons at $n_{x} n_{y}$ lattice cells, respectively. In fact, Eqs.~(\ref{OperFI}a) and (\ref{OperFI}b) correspond to a convenient single (condensate) mode approximation that relates to each site of the lattice in Fig.~\ref{PicLat} \cite{BEC-book}. In particular, the annihilation operators $\hat{a}_{n_{x} n_{y}}$ and $\hat{b}_{n_{x} n_{y} } $ characterize the dynamical properties of atomic ensembles (single atomic quantum modes) at lower (${\left|a\right\rangle}$) and upper (${\left|b\right\rangle}$) levels. The annihilation operator $\hat{\psi}_{n_{x} n_{y}}$ in Eq.(\ref{OperFI}c) describes the temporal behavior of a single photonic mode located at the cavity site.

Thereafter, we restrict ourselves by a tight binding approximation if the coupling between neighbor sites is weak enough~\cite{SmerziTrombettoniKevrekidisBishop}.
Plugging Eq.~(\ref{OperFI}) into Eq.~(\ref{HAPI}) for the parts of Hamiltonian~$\hat{H}$ under the rotating wave approximation one can obtain
\begin{subequations}
\label{HamTBA}
\begin{eqnarray}
\hat{H}_{at}&=&\hbar {\sum _{n_{x}, n_{y}}} \left[\omega_{n_{x} n_{y},\: at}^{(a)} \hat{a}_{n_{x} n_{y}}^{\dag} \hat{a}_{n_{x} n_{y}}+\omega _{n_{x} n_{y},\: at}^{(b)} \hat{b}_{n_{x} n_{y}}^{\dag} \hat{b}_{n_{x} n_{y}} \right. \nonumber\\
&-&\beta_{n_{x}}^{(a)} \left(\hat{a}_{n_{x} n_{y}}^{\dag} \hat{a}_{n_{x}+1\: n_{y}}+\hat{a}_{n_{x} n_{y}}^{\dag} \hat{a}_{n_{x}-1\: n_{y}}\right) \nonumber\\
&-&\beta_{n_{y}}^{(a)} \left(\hat{a}_{n_{x} n_{y}}^{\dag } \hat{a}_{n_{x} n_{y}+1}+\hat{a}_{n_{x} n_{y}}^{\dag} \hat{a}_{n_{x} n_{y}-1} \right) \nonumber\\
&-&\beta_{n_{x}}^{(b)} \left(\hat{b}_{n_{x} n_{y}}^{\dag} \hat{b}_{n_{x}+1 \: n_{y}}+\hat{b}_{n_{x} n_{y}}^{\dag} \hat{b}_{n_{x}-1 \: n_{y}} \right) \nonumber\\
&-& \beta_{n_{y}}^{(b)} \left(\hat{b}_{n_{x} n_{y}}^{\dag} \hat{b}_{n_{x} n_{y}+1}+\hat{b}_{n_{x} n_{y}}^{\dag} \hat{b}_{n_{x} n_{y}-1} \right)\nonumber\\
&+&\frac{1}{2}u_{a} \left(\hat{a}_{n_{x} n_{y}}^{\dag}\right)^{2} \left(\vphantom{\hat{a}_{n_{x} n_{y}}^{\dag}} \hat{a}_{n_{x} n_{y}}\right)^{2} \nonumber\\
&+& \frac{1}{2}u_{b}\left(\hat{b}_{n_{x} n_{y}}^{\dag}\right)^{2} \left(\hat{b}_{n_{x} n_{y}}\right)^{2}\nonumber\\
&+&\left. u_{ab}\ \hat{a}_{n_{x} n_{y}}^{\dag}\hat{a}_{n_{x} n_{y}}\hat{b}_{n_{x} n_{y}}^{\dag}\hat{b}_{n_{x} n_{y}} \right],\label{HamTBAa}
\\
\hat{H}_{ph} &=&\hbar \sum_{n_{x}, n_{y}} \left[\omega_{n_{x} n_{y},\: ph} {\hat{\psi}_{n_{x} n_{y}}}^{\dag} \hat{\psi}_{n_{x} n_{y}}\right. \nonumber\\
&-&\alpha_{n_{x}} \left({\hat{\psi}_{n_{x} n_{y}}}^{\dag} \hat{\psi}_{n_{x}+1 n_{y}}+{\hat{\psi}_{n_{x}n_{y}}}^{\dag} \hat{\psi}_{n_{x}-1 n_{y}} \right)\nonumber\\
& -& \left.\alpha_{n_{y}} \left(\hat{\psi}_{n_{x} n_{y}}^{\dag} \hat{\psi}_{n_{x} n_{y}+1}+\hat{\psi}_{n_{x} n_{y}}^{\dag} \hat{\psi}_{n_{x} n_{y}-1} \right) \right],\label{HamTBAb}
\\
\hat{H}_{int}&=&\hbar \sum_{n_{x} ,n_{y}}g_{n_{x} n_{y}} \left[\hat{\psi}_{n_{x} n_{y}}^{\dag} \hat{a}_{n_{x} n_{y}}^{\dag} \hat{b}_{n_{x} n_{y}}\right.\nonumber\\
&+& \left.\hat{b}_{n_{x} n_{y}}^{\dag} \hat{a}_{n_{x} n_{y}} \hat{\psi}_{n_{x} n_{y}} \right],\label{HamTBAc}
\end{eqnarray}
\end{subequations}
where $g_{n_{x} n_{y}} =\kappa \int \xi _{n_{x} n_{y}} \left(\mathbf{r}\right)\varphi _{n_{x} n_{y}}^{(a)} \left(\mathbf{r}\right)\varphi _{n_{x} n_{y}}^{(b)} \left(\mathbf{r}\right)d^{3}{\mathbf{r}}$; $\omega_{n_{x} n_{y},\:  at}^{(a,b)}$ and $\omega_{n_{x} n_{y},\:  ph}$ are the frequencies for atoms and photons in the lattice, respectively; and $u_{a,b} =\frac{1}{\hbar } U_{a,b} \int \left(\varphi _{n_{x} n_{y} }^{(a,b)} \right)^{4} d^{3} \mathbf{r} $ and $u_{ab} =\frac{1}{\hbar } U_{ab} \int \left(\varphi _{n_{x} n_{y} }^{(a)} \right)^{2} \left(\varphi _{n_{x} n_{y} }^{(b)} \right)^{2} d^{3} \mathbf{r}$ characterize  frequencies of nonlinear self and cross  atomic mode interactions.
Hopping constants $\beta_{n_{x}, n_{y}}^{(a,b)}$ are calculated by performing an integral for atomic wave function overlapping in the adjacent sites, i.e., nearest-neighbor hopping constants for the atoms in the lattice, while $\alpha_{n_{x}, n_{y}}$ characterize a spatial overlapping of optical fields between the neighboring cavities.

Let us discuss the main approximations and bounds for the applicability of the PolC model~in~Fig.~\ref{PicLat}. 

First, since we are interested in the mean-field properties of a polaritonic system, we assume for simplicity that all cavities are identical to each other and contain the same average number of atoms $N =  \langle \hat{a}_{n_{x} n_{y}}^{\dag} \hat{a}_{n_{x} n_{y}} +\hat{b}_{n_{x} n_{y}}^{\dag} \hat{b}_{n_{x} n_{y}}\rangle$.  Furthermore, we suppose that atom-light coupling coefficients are equal to each other at all sites by assuming $g_{0} \equiv g_{n_{x} n_{y}} $.

Second, we use a single mode approach for atomic ensembles assuming that a motional degree of freedom is unimportant.  It is possible to show that this approximation  is valid  only when the size of the atomic trap (cavity size) is much larger than the parameter $N\left|a_{j}^{(sc)} \right|, (j=a,b,ab)$  at each site of the lattice~(cf.~\cite{AnglinVardiPRA}). Taking into account a typical value of atomic scattering length, $\left|a_{j}^{(sc)} \right|\simeq 5\text{ nm}$, and a maximal cavity size, $\ell \simeq 3\;\mu \text{m}$, it is possible to estimate a maximally available  total  number of atoms as $N\simeq 800$ for each site.

Third, we are working under the  strong atom-field coupling condition for which the inequality 
\begin{equation}
\label{NonewqualityG}
g_{0} >\Gamma ,\gamma _{ph}
\end{equation} 
is satisfied; $\Gamma $ and $\gamma _{ph} $ are the spontaneous emission and cavity field decay rates, respectively.  To be more specific, we consider that a quantized optical field interacts with ensembles of two-level rubidium atoms, which have resonance frequency $\omega_{ab} \mathord{\left/ {\vphantom {\omega_{ab} 2\pi}} \right. 2\pi }=382 \text{ THz}$ corresponding to rubidium $D$ lines~\cite{ChestnovAlodjantsArakelian}. The lifetime for rubidium atoms is taken as $27 \text{ ns}$ corresponding to the spontaneous emission rate $\Gamma $ of about $2\pi \times 6\text{ MHz}$.

A cavity field decay rate is defined as $\gamma _{ph} =\omega _{c} /2 Q$, where $\omega _{c} $ is the frequency of the cavity mode and  $Q$ is the cavity quality factor. At present it is practically possible to achieve the values of  $Q\simeq 10^{5} \div 10^{6}$  for photonic crystal microcavities~(cf.~\cite{VahalaNature003}), which implies, for example, the value ${\gamma _{ph} / 2 \pi \simeq 0.955\text{ GHz}}$ for the cavity decay rate, taken at atom-field resonance for $Q\simeq 2 \times 10^{5}$. 

The strength of interaction of a single atom with a quantum optical field is taken as ${g_{0} =\left({\left|d_{ab} \right|^{2} \omega_{ab} \mathord{\left/ {\vphantom {\left|d_{ab} \right|^{2}}} \right.} 2\hbar \varepsilon_{0} V_{M}} \right)}^{1/2}$ at each cavity with the atomic dipole matrix element $d_{ab} $ and the interaction (mode) volume $V_{M} $. To achieve a strong atom-field coupling regime \eqref{NonewqualityG} the mode volume $V_{M} $ has to be as small as possible. Practically it is possible to reach  $V_{M} \simeq \left(\lambda_{0} /2\right)^{3} $, where  $\lambda_{0} $ is a light-field wavelength~(cf~\cite{LeeYariv,VahalaNature003}). In this case the atom-field coupling  strength $g_{0} /2\pi $ is of the value of a few gigahertz.

Next, we follow the Holstein-Primakoff transformation by mapping atomic excitation operators~$\hat{\phi}_{n}$ and $\hat{\phi}_{n}^{\dag}$ into a Schwinger representation for a two-level oscillator system~\cite{HP}, i.e., 
\begin{subequations}
\label{OHP}
\begin{eqnarray}
\hat{S}_{+, n} &=&\hat{\phi}_{n}^{\dag} \sqrt{N-\hat{\phi}_{n}^{\dag} \hat{\phi}_{n}} ,\label{OHPa}
\\
\hat{S}_{-, n} &=&\left(\sqrt{N-\hat{\phi}_{n}^{\dag} \hat{\phi}_{n}}\right)\:\hat{\phi}_{n},\label{OHPb}
\\
\hat{S}_{z, n} &=&\hat{\phi}_{n}^{\dag} \hat{\phi}_{n} -{N\mathord{\left/2\right.}},\label{OHPc}
\end{eqnarray}
\end{subequations}
where the operators are defined as ${\hat{S}_{+, n}=\hat{b}_{n}^{\dag} \hat{a}_{n}}$, ${\hat{S}_{-, n}=\hat{a}_{n}^{\dag} \hat{b}_{n}}$, and ${\hat{S}_{z, n}=\frac{1}{2} \left(\hat{b}_{n}^{\dag} \hat{b}_{n}-\hat{a}_{n}^{\dag} \hat{a}_{n}\right)}$, 
and ${n\equiv \left\{n_{x}, n_{y} \right\}}$.
If~${\hat{\phi} _{n}^{\dag} \hat{\phi}_{n} \simeq \hat{b}_{n}^{\dag} \hat{b}_{n}}$ approximates the atomic excitations, then it is possible to treat operators describing atoms at lower and upper levels as 
\begin{subequations}
\label{Ops}
\begin{eqnarray}
\hat{b}_{n} &\simeq & \hat{\phi}_{n},\label{Opsa}\\
\hat{b}_{n}^{\dag} &\simeq& \hat{\phi}_{n}^{\dag} ,\label{Opsb}
\end{eqnarray}
\begin{eqnarray}
\hat{a}_{n}, \hat{a}_{n}^{\dag} &\simeq & \sqrt{N-\hat{\phi}_{n}^{\dag} \hat{\phi}_{n}} \nonumber\\
&\approx& \sqrt{N} -\frac{\hat{\phi}_{n}^{\dag } \hat{\phi}_{n}}{2N^{1/2}} -\frac{\left(\hat{\phi}_{n}^{\dag} \hat{\phi}_{n} \right)^{2}}{8N^{3/2}}.\label{Opsc}
\end{eqnarray}
\end{subequations}

Evidently, such an approximation is only valid for a macroscopic number of atoms being at \textit{coherent state} at each cell of the lattice when relative \textit{quantum phase} properties of the atoms at the ground state for neighbor sites can be ignored.
Traditionally, the Bogoliubov approach to studying superfluidity is restricted by keeping the first term in Eq. (\ref{Opsc}) for the expansion of atomic operators~$\hat{a}_{n}$ and $\hat{a}_{n}^{\dag}$, respectively (see, e.g., \cite{BEC-book}). It leads to the so-called low-excitation-density limit, i.e., 
$\left\langle \hat{b}_{n}^{\dag } \hat{b}_{n} \right\rangle \ll \left\langle \hat{a}_{n}^{\dag } \hat{a}_{n} \right\rangle$~(cf. \cite{AlodjantsBarinovArakelian}).
The low excitation limit implies that the atoms mostly populate their ground level ${\left| a \right\rangle}$. It can also be rewritten in a slightly different form as $\left\langle \hat{\phi}_{n}^{\dag} \hat{\phi}_{n} \right\rangle \ll N \mathord{\left/ {\vphantom {N 2}} \right. 2}$, which is typically considered in the framework of exciton-polariton BEC analysis ~\cite{DengHaugYamamoto,KasprzakRichardKundermann,BaliliHartwellSnoke,UtsunomiyaTianRoumpos}. 

In this paper we keep all terms in the expansion of~Eq.~(\ref{Opsc}). In this limit operators $\hat{a}_{n}$ and $\hat{a}_{n}^{\dag}$ represent \textit{q}-deformed bosonic operators and characterize saturation effects occurring beyond the low-density limit~(cf.~\cite{SteynRossGardinerPRA}). Combining Eqs.~(\ref{Ops}) and Eqs.~(\ref{HamTBA}a) and~(\ref{HamTBA}c), we rewrite the Hamiltonians $\hat{H}_{at} $ and $\hat{H}_{int} $ containing atomic operators as
\begin{subequations}
\label{HamKoefIOp}
\begin{eqnarray}
\hat{H}_{at}&=&\hbar \sum_{n_{x}, n_{y}} \left[ \vphantom {\frac{\hbar}{2}}  \left(\omega _{n_{x} n_{y},\: at}^{(b)}-\omega _{n_{x} n_{y},\: at}^{(a)} \right. \right. \nonumber\\
&+& \left. 2\beta _{n_{x}}^{(a)}+2\beta _{n_{y}}^{(a)} +\left( u_{ab} -u_a \right)N\right)\hat{\phi}_{n_{x} n_{y}}^{\dag } \hat{\phi}_{n_{x} n_{y}} \nonumber \\
&-& \beta _{n_{x}}^{(b)} \left(\hat{\phi}_{n_{x} n_{y}}^{\dag } \hat{\phi}_{n_{x} +1\: n_{y}} +\text{H.c.}\right) \nonumber\\
&-&\beta _{n_{y}}^{(b)} \left(\hat{\phi}_{n_{x} n_{y}}^{\dag } \hat{\phi}_{n_{x} n_{y}+1}+\text{H.c.}\right)  \nonumber\\
&+&\left.\frac{ u}{2} \left( \hat{\phi}_{n_{x} n_{y}}^{\dag} \right)^{2} \left(\hat{\phi}_{n_{x} n_{y}}\right)^{2} \right],\label{HamKoefIOpa}
\\
\hat{H}_{int} &=&
\hbar g\sum _{n_{x}, n_{y}} \left[ \hat{\psi}_{n_{x} n_{y}}^{\dag} \hat{\phi}_{n_{x} n_{y}}+\text{H.c.}\right]\nonumber\\
&-&\frac{\hbar g}{2N} \sum _{n_{x}, n_{y}}\left[\hat{\psi} _{n_{x} n_{y}}^{\dag} \hat{\phi}_{n_{x} n_{y}}^{\dag} \left( \hat{\phi}_{n_{x} n_{y}}\right)^{2} \right. \nonumber\\
&+&\left.\frac{1}{4N} \hat{\psi}_{n_{x} n_{y}}^{\dag} \left( \hat{\phi}_{n_{x} n_{y}}^{\dag}\right)^{2} \left(\hat{\phi}_{n_{x} n_{y}}\right)^{3}+\text{H.c.}\right], \qquad \label{HamKoefIOpc}
\end{eqnarray}
\end{subequations}
where $g=g_{0}\sqrt{N}$ is a collective atom-field coupling constant taken at each site of the lattice; $u=u_{a}+u_{b}-2u_{ab}+\frac{1}{N}\beta_{n_{x}}^{(a)}+\frac{1}{N}\beta_{n_{y}}^{(a)}$ characterizes nonlinear effects in excitations of a two-level atomic system.

Then we take the $\mathbf{k}$ representation for the Hamiltonians in Eqs.~(\ref{HamTBAb}),~(\ref{HamKoefIOp}a), and~(\ref{HamKoefIOp}b) relying on  the periodical properties of our PolC system and introduce the operators $\hat{\phi}_{n}$~$\equiv$~$\hat{\phi}_{n_{x} n_{y}}$ and $\hat{\psi}_{n}$~$\equiv$~$\hat{\psi}_{n_{x} n_{y}}$ in the form of
\begin{subequations}
\label{PhiInK}
\begin{eqnarray}
\label{PhiInKa}
\hat{\phi}_{n} &=& \frac{1}{\sqrt{M}} \sum_{\mathbf{k}}\hat{\phi}_{\mathbf{k}} {e}^{i \mathbf{k} \bm{\ell }},\\
\label{PhiInKb}
\hat{\psi}_{n} &=& \frac{1}{\sqrt{M}} \sum _{\mathbf{k}}\hat{\psi}_{\mathbf{k}} e^{i \mathbf{k} \bm{\ell }}, 
\end{eqnarray}
\end{subequations}
where $\bm{\ell }$ is a lattice vector.
For a 2D periodic structure of PolC we have $\mathbf{k} \bm{\ell }=k_{x} n_{x} \ell _{x}+$ $k_{y} n_{y} \ell _{y}$, $n_{x}=1,2,...,N_{x} $, $n_{y} =1,2,...,N_{y} $, and $M=N_{x} \times N_{y}$,  with the lattice constants $\ell_{x}$, $\ell _{y}$ in $x$ and $y$ directions, respectively. For anisotropic lattice configuration we may have $\ell_{x} \ne \ell_{y}$. By substituting Eqs.~(\ref{PhiInK}a) and~(\ref{PhiInK}b) for Eqs.~(\ref{HamTBAb}),~(\ref{HamKoefIOp}a) and~(\ref{HamKoefIOp}b), we arrive at a $\mathbf{k}$-space expression for the Hamiltonian, which can be written as a linear one and a nonlinear one, i.e., 
\begin{subequations}
\label{HamInK}
\begin{eqnarray}
&&\hat{H}=\hat{H}^{(L)} +\hat{H}^{(NL)},\label{HamInKa}
\\
&&\hat{H}^{(L)}=\hbar \sum_{\mathbf{k}}\left[\omega_{at} \hat{\phi}_{\mathbf{k}}^{\dag} \hat{\phi}_{\mathbf{k}} +\omega _{ph} \hat{\psi}_{\mathbf{k}}^{\dag} \hat{\psi}_{\mathbf{k}} \right.\nonumber\\
&&\phantom{\hat{H}^{(L)}} \left. +g \left(\hat{\psi}_{\mathbf{k}}^{\dag} \hat{\phi}_{\mathbf{k}} +\hat{\phi}_{\mathbf{k}}^{\dag} \hat{\psi}_{\mathbf{k}} \right)\right], \label{HamInKb}
\\
&&\hat{H}^{(NL)}=\frac{\hbar u}{2M}\sum_{\mathbf{k}_{1,2}\, ,\mathbf{q}} \hat{\phi}_{\mathbf{k}_{1} +\mathbf{q}}^{\dag} \hat{\phi}_{\mathbf{k}_{2} -\mathbf{q}}^{\dag} \hat{\phi}_{\mathbf{k}_{2}} \hat{\phi} _{\mathbf{k}_{1}} \nonumber\\ 
&&\phantom{\hat{H}^{(NL)}}-\frac{\hbar g}{2N_{tot} }\sum_{\mathbf{k}_{1,2}\, ,\mathbf{q}} \left[\hat{\psi}_{\mathbf{k}_{1} +\mathbf{q}}^{\dag} \hat{\phi}_{\mathbf{k}_{2} -\mathbf{q}}^{\dag} \hat{\phi}_{\mathbf{k}_{2}} \hat{\phi} _{\mathbf{k}_{1}} +\text{H.c.}\right] \nonumber\\ 
&&\phantom{\hat{H}^{(NL)}} -\frac{\hbar g}{8N_{tot}^{2}}\sum_{\mathbf{k},\, \mathbf{k}_{1,2},\, \mathbf{q}_{1,2}}  \left[\hat{\psi}_{\mathbf{k}+\mathbf{q}_{1} + \mathbf{q}_{2}}^{\dag} \hat{\phi}_{\mathbf{k}_{1} - \mathbf{q}_{1}}^{\dag}\hat{\phi}_{\mathbf{k}_{2} - \mathbf{q}_{2}}^{\dag} \right.\nonumber\\
&&\phantom{\hat{H}^{(NL)}} \times\left. \hat{\phi}_{\mathbf{k}_{2}} \hat{\phi}_{\mathbf{k}_{1}} \hat{\phi}_{\mathbf{k}} +\text{H.c.}\right],\label{HamInKc}
\end{eqnarray}
\end{subequations}
where $N_{tot}=NM$ is the total number of atoms at all sites. The frequencies $\omega _{at} (k)\equiv \omega _{at}^{\left(b\right)} -\omega _{at}^{\left(a\right)}$ and $\omega _{ph} (k)$ characterize the dispersion properties of atomic and photonic states in a PolC structure, which are determined by
\begin{subequations}
\label{DispRel}
\begin{eqnarray}
\omega_{at}&=& \omega_{n_{x} n_{y},\: at}^{\left(b\right)}-\omega _{n_{x} n_{y},\: at}^{\left(a\right)} \nonumber\\
&+& 2\sum_{j=x,y} \left(\beta_{n_{j}}^{(a)} -\beta _{n_{j}}^{(b)} \cos k_{j} \ell _{j} \right)+\left( u_{ab}-u \right)N,  \qquad\,\, \label{DispRela}\\
\omega_{ph}&=&\omega_{n_{x} n_{y},\: ph} -2\sum_{j=x,y}\alpha _{n_{j}} \cos k_{j} \ell _{j} \label{DispRelb} .
\end{eqnarray}
\end{subequations}

For small quasimomentum components, these dispersion relations can be approximated as 
\begin{subequations}
\label{DispRelSmall}
\begin{eqnarray}
&&\omega _{at}(k)\simeq \omega _{at}^{\left(ba\right)}+\sum_{j=x,y}\frac{\hbar k_{j}^{2}}{2m_{at,\:j}}, \label{DispRelSmalla}
\\
&&\omega _{ph} (k)\simeq \omega _{L} +\sum_{j=x,y} \frac{\hbar k_{j}^{2}}{2m_{ph,\:j}}, \label{DispRelSmallb}
\end{eqnarray}
\end{subequations}
where the related atomic and photonic frequencies $\omega_{at} (k)$ and $\omega_{ph} (k)$ are taken at the center of Brillouin zone, i.e., $\omega_{at}^{\left(ba\right)} \equiv \left. \omega_{at} (k)\right|_{k=0}, \; \omega_{L}=\left. \omega_{ph}(k)\right|_{k=0}$. In Eqs.~(\ref{DispRelSmall}) we also introduce effective lattice masses for photons and atoms, denoted by ${m_{ph,\:j}=\hbar/2\alpha_{n_{j}} \ell _{j}^{2}}$ and ${m_{at,\:j} = \hbar/{2\beta_{n_{j}}^{(b)} \ell_{j}^{2}}}$ ${(j=x,y)}$, respectively.

Hamiltonian $\hat{H}$ in Eqs.~(\ref{HamInK}) represents a \textit{many-body} Hamiltonian describing the atom-field interaction in the momentum space for  a 2D PolC structure. 
The linear part $\hat{H}^{(L)}$ of this Hamiltonian is usually examined in the framework of upper- and lower-branch polaritons~\cite{DengHaugYamamoto}.
But the nonlinear part $\hat{H}^{(NL)}$ characterizes polariton interaction effects beyond the low-density limit.
The latter one is the main result of the present work and is used to study the nonlinear dynamics of PolC in the following sections.

\section{\label{polaritonpropertiesinmomentum}POLARITON PROPERTIES IN MOMENTUM REPRESENTATION}
If a quantum field intensity in the lattice is not too high, we can assume that the corresponding dispersion relation for a polariton states is not modified.
Thus, we use the polariton basis to diagonalize the total Hamiltonian. In particular, one can use the following linear transformations to couple cavity photons and atomic excitations, 
\begin{subequations}
\label{PolOps}
\begin{eqnarray}
\hat{\Xi} _{1,\mathbf{k}} &=&X_{\mathbf{k}} \hat{\psi} _{\mathbf{k}} +C_{\mathbf{k}} \hat{\phi} _{\mathbf{k}},\label{PolOpsa} \\
\hat{\Xi} _{2,\mathbf{k}} &=& X_{\mathbf{k}} \hat{\phi} _{\mathbf{k}} -C_{\mathbf{k}} \hat{\psi} _{\mathbf{k}},\label{PolOpsb}
\end{eqnarray}
\end{subequations}
where $X_{\mathbf{k}}$ and $C_{\mathbf{k}}$ are the corresponding Hopfield coefficients defined as
\begin{subequations}
\label{HopKoeff}
\begin{eqnarray}
X_{\mathbf{k}} &=&\frac{1}{\sqrt{2} } \left(1+\frac{\delta _{\mathbf{k}} }{\sqrt{4g^{2} +\delta _{\mathbf{k}}^{2} } } \right)^{1/2},\label{HopKoeffa}\\
C_{\mathbf{k}} &=&\frac{1}{\sqrt{2} } \left(1-\frac{\delta _{\mathbf{k}} }{\sqrt{4g^{2} +\delta _{\mathbf{k}}^{2} } } \right)^{1/2} . \label{HopKoeffb}
\end{eqnarray}
\end{subequations}

Here we have defined a quasimomentum-dependent frequency mismatch 
\begin{eqnarray}
\delta_{\mathbf{k}} &\equiv& \omega_{ph} (k)-\omega _{at} (k)\nonumber\\
&=& \Delta +2\sum_{j=x,y}\left(\beta_{n_{j}}^{(b)} -\alpha_{n_{j}} \right)\cos k_{j} \ell_{j},
\end{eqnarray}
with \textit{momentum-independent} atom-field detuning $\Delta=\omega_{n_{x} n_{y},\: ph} -\left[\omega_{n_{x} n_{y},\: at}^{\left(b\right)} -\omega_{n_{x} n_{y},\: at}^{\left(a\right)}+2\left(\beta_{n_{x}}^{(a)} +\beta_{n_{y}}^{(a)} \right)+\left( u_{ab} -u \right)N\right]$.
%$\Delta=\omega_{n_{x} n_{y},\: ph} -\left[\omega_{n_{x} n_{y},\: at}^{\left(b\right)} -\omega_{n_{x} n_{y},\: at}^{\left(a\right)}+2\left(\beta_{n_{x}}^{(a)} +\beta_{n_{y}}^{(a)} \right)\right]$
Parameters $X_{\mathbf{k}} $ and $C_{\mathbf{k}}$ are symmetric and normalized with respect to quasimomentum, i.e., $X_{\mathbf{k}} = X_{-\mathbf{k}}$ , $C_{\mathbf{k}} = C_{-\mathbf{k}}$, and $C_{\mathbf{k}}^{2} + X_{\mathbf{k}}^{2} = 1$. 

\begin{figure}
\includegraphics[width=7.5cm]{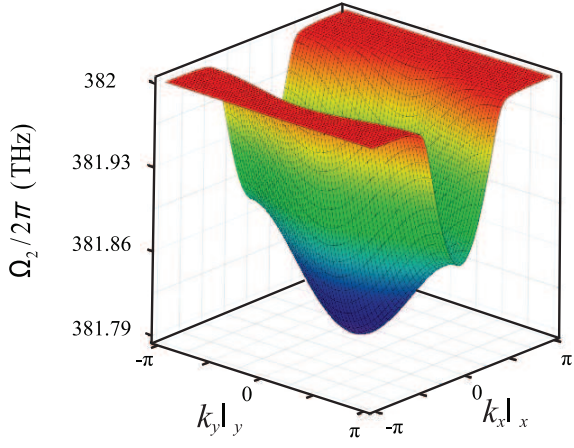}%for PDFLaTeX
\caption{\label{PicOm3D} (Color online) Dependence of the characteristic frequency $\Omega_{2} \left(k_{x}, k_{y} \right) /2 \pi$ for LB polaritons on the quasimomentum components in the first Brillouin zone. The parameters used are the following: the average number of rubidium atoms at each cavity is taken as $N=100$; the collective atom-field coupling strength is  $g/2\pi=12.2 \text{ GHz}$; the total number of cavities is $M=100$; the photon masses in the lattice are $m_{ph,x} \simeq m_{ph,y} =2.8\times 10^{-36} \text{ kg}$; the ratio of lattice constants is ${\alpha _{n_{x}} / \alpha _{n_{y}}}=4$ for $\ell_{x} =6\,\mu\text{m}$ and $\ell_{y} =3\,\mu\text{m}$; and the atom-field detuning is $\Delta =0$.}

\end{figure}
Operators $\hat{\Xi}_{1,\mathbf{k}}$ and $\hat{\Xi}_{2,\mathbf{k}}$ characterize two types of bosonic quasiparticles under the atom-field interaction, i.e., upper- and lower-branch polaritons. 
At the low-density limit, these two branches of polariton states are the exact solutions of the linear Hamiltonian $H^{(L)}$, with two characteristic frequencies $\Omega_{1,2} (k)$ defined by 
\begin{equation}
\label{PolDispRel}
\Omega _{1,2} \left(k\right)=\frac{1}{2} \left(\omega _{at} (k)+\omega _{ph} (k)\pm \sqrt{4g^{2}+\delta _{\mathbf{k}}^{2}} \right),
\end{equation}
determining a dispersion relation for polaritons in a band-gap structure. In Fig.~\ref{PicOm3D} the dispersion relation for LB polaritons is examined in the first Brillouin zone of a periodic structure.
The principal feature of the dispersion surface shown in Fig.~\ref{PicOm3D} is the presence of the energy minimum for polaritons at $k_{x} =k_{y} =0$. A flat region on this surface appears due to a small Rabi splitting frequency in comparison with an atomic transition frequency, i.e., $g \ll \omega_{ab}$.
Following this peculiarity, one can approximate the dispersion relation for LB polaritons as a parabolic curve,
\begin{equation}
\label{DispParab}
\Omega _{\mathbf{k}} \equiv \Omega _{2} \left(k_{x}, k_{y}\right)\simeq \frac{\hbar k_{x}^{2}}{2m_{x}} +\frac{\hbar k_{y}^{2}}{2m_{y}},
\end{equation} 
which is relevant for small quasimomentum components. In Eq.~(\ref{DispParab}) we introduce the LB polariton mass $m_{x,y}$ in spatial directions.
Thus, polaritons in the lattice structure are represented as massive particles in two spatial directions with tensorial mass $m_{x}$ and $m_{y}$~\cite{FleischhauerOtterbachUnanyan}.

This parabolic-type dispersion relation for $\Omega _{2} \left(k\right)$ describes free quasiparticles (polaritons) at the bottom of the dispersion surface in Fig.~\ref{PicOm3D}. 
Such a characteristic of the LB atomic polariton dispersion can be used to achieve a BEC state with a quasimomentum $\mathbf{k}=0$~\cite{AlodjantsBarinovArakelian}.
It is well known that such a phase transition for a uniform 2D gas of Bose particles occurs at temperature $T=0$~\cite{BEC-book}. 
But a finite (non-zero) temperature of polariton condensation can only be achieved by trapping LB polaritons~\cite{BaliliHartwellSnoke}. 
An interesting feature of polariton BEC is that its corresponding critical temperature $T_{C} \propto {1\mathord{\left/ {\vphantom {1 \sqrt{m}}} \right.} \sqrt{m}}$ can be many orders higher than that of an atomic ensemble due to a small polariton mass $m$.
For example, for isotropic (symmetric) lattices, the polariton mass is $m=2m_{ph} \simeq 5.6\times 10^{-36}\text{ kg}$ under the atom-field resonance condition $\tilde{\Delta}=0$. However, critical features of LB polaritons in PolC are limited by the temperature of maintaining coherent properties of a 2D combined atom-light structure presented in Fig. \ref{PicLat}.

The interaction between polaritons occurs due to the nonlinear term $\hat{H}^{(NL)} $ in Eq. (\ref{HamInKc}). 
By keeping LB polariton terms, the total Hamiltonian $\hat{H}$ can be restored as
\begin{eqnarray}
\label{HamPolSum}
\hat{H}&=&\sum _{\mathbf{k}}\hbar \Omega _{k} \hat{\Xi} _{\mathbf{k}}^{\dag } \hat{\Xi} _{\mathbf{k}} \nonumber\\
&+&\frac{1}{2M} \sum _{\mathbf{k}_{1,2}, \mathbf{q}}U_{\mathbf{k}_{1}\mathbf{k}_{2} \mathbf{q}}^{(1)} \hat{\Xi} _{\mathbf{k}_{1}+\mathbf{q}}^{\dag} \hat{\Xi} _{\mathbf{k}_{2}-\mathbf{q}}^{\dag} \hat{\Xi} _{\mathbf{k}_{2}} \hat{\Xi} _{\mathbf{k}_{1}} \nonumber\\
&+&\frac{1}{2M} \sum _{\mathbf{k},\mathbf{k}_{1,2}, \mathbf{q}_{1,2}}U_{\mathbf{k}\mathbf{k}_{1} \mathbf{k}_{2} \mathbf{q}_{1} \mathbf{q}_{2}}^{(2)} \hat{\Xi} _{\mathbf{k}+\mathbf{q}_{1}+\mathbf{q}_{2}}^{\dag} \hat{\Xi} _{\mathbf{k}_{1}-\mathbf{q}_{1}}^{\dag } \hat{\Xi} _{\mathbf{k}_{2}-\mathbf{q}_{2}}^{\dag} \nonumber\\
&\times & \hat{\Xi} _{\mathbf{k} _{2}} \hat{\Xi} _{\mathbf{k} _{1}} \hat{\Xi} _{\mathbf{k}},
\end{eqnarray}
where we omit index label ``2'' at LB polariton operators for simplicity. In Eq. (\ref{HamPolSum}) we also introduce two polariton interaction parameters,
\begin{subequations}
\begin{eqnarray}
&&U_{\mathbf{k}_{1} \mathbf{k}_{2} \mathbf{q}}^{(1)} =\hbar \left[  u X_{\left|\mathbf{k}_{1}+\mathbf{q}\right|} X_{\mathbf{k}_{2}}  + \frac{ g}{N} \left(C_{\left|\mathbf{k}_{1}+\mathbf{q}\right|} X_{\mathbf{k}_{2}}\right. \right. \nonumber\\
&&\phantom{U_{\mathbf{k}_{1} \mathbf{k}_{2} \mathbf{q}}^{(1)}} \left. \left. + C_{\mathbf{k}_{2}} X_{\left|\mathbf{k}_{1}+\mathbf{q}\right|} \right)\right] X_{\left|\mathbf{k}_{2}-\mathbf{q}\right|} X_{\mathbf{k}_{1} } ,\\
&&U_{\mathbf{k}\mathbf{k}_{1} \mathbf{k}_{2} \mathbf{q}_{1} \mathbf{q}_{2}}^{(2)}=
\frac{\hbar g}{4MN^{2}}\left(C_{\left|\mathbf{k}+\mathbf{q}_{1}+\mathbf{q}_{2} \right|} X_{\mathbf{k}_{1}} \right. \nonumber\\
&&\phantom{U_{\mathbf{k}\mathbf{k}_{1}}} \left. +C_{\mathbf{k}_{1}} X_{\left|\mathbf{k}+\mathbf{q}_{1}+\mathbf{q}_{2} \right|} \right)X_{\left|\mathbf{k}_{2}-\mathbf{q}_{2} \right|} X_{\left|\mathbf{k}_{1}-\mathbf{q}_{1} \right|} X_{\mathbf{k}_{2}} X_{\mathbf{k}},\qquad
\end{eqnarray}
\end{subequations}
which are relevant to two- and three-body polariton-polariton collisions, respectively.

\begin{figure}
\includegraphics[width=7.5cm]{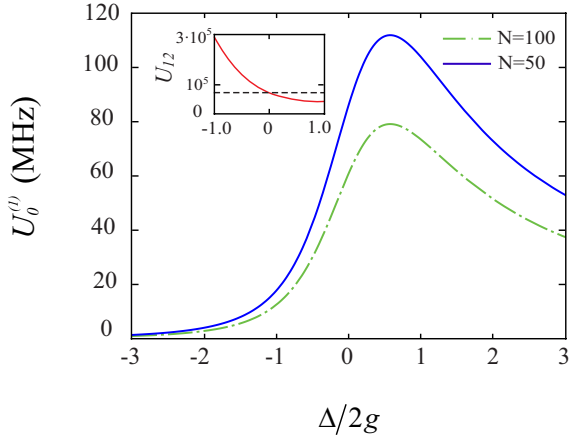}%for PDFLaTeX
\caption{\label{PicU} (Color online) Strength of polariton-polariton interaction $U_{0}^{(1)} \equiv U_{\mathbf{k}}^{(1)} / 2\pi \hbar $ as a function of normalized atom-field detuning ${\Delta / 2g} $ at $\mathbf{k}=0$. The lattice constants are ${\ell_{x}=\ell_{y}=3\mu\text{m}}$. Other parameters are the same as those in Fig.~\ref{PicOm3D}. The inset shows the ratio $U_{12} \equiv U_{0}^{(1)} / U_{0}^{(2)} $ between two- and three-body polariton scatterings. A dashed line corresponds to the case of a half-matter half-photon polariton with a zero detuning of $\Delta =0$. }
\end{figure}

In Eq.~(\ref{HamPolSum}) we ignore the terms which describe interactions between lower and upper polariton branches. It seems to be justified if condition $k_{B}T \ll \hbar g$ is fulfilled. For example, in the course of current experiments with exciton polaritons in semiconductor microstructures the low branch of polaritons is essentially more populated at thermal equilibrium~(see, e.g.,~\cite{KasprzakRichardKundermann,BaliliHartwellSnoke,UtsunomiyaTianRoumpos}).
For a sufficiently low temperature, we can also take polariton scattering parameters close to the zero quasimomentum by assuming
\begin{subequations}
\label{ParScattZero}
\begin{eqnarray}
U_{0}^{(1)} &\equiv& \left. U_{\mathbf{k}_{1} \mathbf{k}_{2} \mathbf{q}}^{(1)} \right|_{\mathbf{k}_{1} ,\mathbf{k}_{2} ,\mathbf{q}=0} =\frac{2\hbar gC_{0} X_{0}^{3} }{N} + \hbar u X_{0}^{4},\qquad \\
 U_{0}^{(2)} &\equiv& \left. U_{\mathbf{k}\mathbf{k}_{1} \mathbf{k}_{2} \mathbf{q}_{1} \mathbf{q}_{2} }^{(2)} \right|_{\mathbf{k},\mathbf{k}_{1} ,\mathbf{k}_{2} ,\mathbf{q}_{1} ,\mathbf{q}_{2}=0} =\frac{\hbar gC_{0} X_{0}^{5} }{2MN^{2} }.
\end{eqnarray}
\end{subequations}

Physically, two nonlinear processes, i.e., the process of atomic collisions and the process of  saturation of two-level atomic systems, contribute to the parameter $U_{0}^{(1)} $ that describes two-body  polariton-polariton scattering (cf.~\cite{KarrBaasGiacobinoPRA}). However, for moderate average atom number  $N<800$ taken at each site of the lattice  the parameter $u n_{pol} $ for trapped rubidium atoms is a few hertz (cf.~\cite{LeggettRMPh}), which is  smaller by many orders than the reduced collective atom-field coupling  strength $g n_{pol} /N$, where $n_{pol} $ is the number of polaritons that can be estimated as the average number of excited atoms at each cavity. In this paper we consider polaritons for which Hopfield coefficients are of the same order, i.e., $X_{\mathbf{k}} \sim C_{\mathbf{k}} $, which corresponds to a moderate atom-field detuning $\Delta$ that is of the order of the atom-field coupling parameter $g$.  In this case a contribution of the last term in Eq.~(\ref{ParScattZero}a) is negligibly  small and we can take  $U_{0}^{(1)} \simeq {2\hbar gC_{0} X_{0}^{3} \mathord{\left/ {\vphantom {2\hbar gC_{0} X_{0}^{3}  N}} \right. \kern-\nulldelimiterspace} N} $ for further processing~(cf.~\cite{KarrBaasGiacobinoPRA}).

Figure~\ref{PicU} demonstrates the behavior of parameter $U_{0}^{(1)}$ as a function of reduced atom-field detuning $\Delta$ taken at the bottom of the dispersion surface. The $U_{0}^{(1)}$ parameter vanishes for a negative detuning ($\Delta <0$) where polaritons become more photonlike. The maximal value of the polariton scattering parameter is achieved for atomlike polaritons with a positive atom-field detuning ${\Delta ={2g\mathord{\left/ {\vphantom {2g \sqrt{3} }} \right. \kern-\nulldelimiterspace} \sqrt{3}}}$. 
 The dependence of the ratio ${U_{12} \equiv {U_{0}^{(1)} \mathord{\left/ {\vphantom {U_{0}^{(1)} U_{0}^{(2)}}} \right.} U_{0}^{(2)}}={4N_{tot} \mathord{\left/ {\vphantom {4N_{tot} X_{0}^{2}}} \right.} X_{0}^{2}}}$ of polariton nonlinear interaction parameters is also outlined in the inset of Fig.~\ref{PicU}.
 It is worth mentioning that when $U_{0}^{(1)} \gg U_{0}^{(2)}$ the last term in Eq. (\ref{HamPolSum}) can be neglected for a negative detuning ($\Delta <0$) for a photonlike polariton.

\section{\label{nonlineardynamicsofpolaritons}NONLINEAR DYNAMICS OF PolC}

\subsection{\label{variationalapproachforpolariton}Variational approach for PolC}
Let us consider the properties of PolC\, in the continuum limit. By treating a many-body Hamiltonian in Eq.~(\ref{HamPolSum}) in the coordinate representation, one can get 

\begin{eqnarray}
\label{HamMain}
\hat{H}&=&\int \left\{\hat{\Psi}^{\dag} (\mathbf{r})\left[-\frac{\hbar ^{2} \partial ^{2} }{2m_{x} \partial x^{2}}-\frac{\hbar ^{2} \partial ^{2} }{2m_{y} \partial y^{2}}+V_{tr}^{(pol)} (\mathbf{r})\right]\hat{\Psi }(\mathbf{r})\right.\nonumber\\
&+&\frac{U_{1}}{2} \hat{\Psi}^{\dag} (\mathbf{r})^{2} \hat{\Psi}(\mathbf{r})^{2} {\left.+\frac{U_{2}}{3} \hat{\Psi}^{\dag } (\mathbf{r})^{3} \hat{\Psi }(\mathbf{r})^{3} \right\}d^{2} \mathbf{r}},
\end{eqnarray} 
where $\hat{\Psi}$ $\left(\hat{\Psi}^{\dag} \right)$ is polariton field annihilation (creation) operator that describes quantum macroscopic properties of PolC.
Related two- and three-body polariton-polariton interaction strengths are defined as $U_{1} =\frac{2\hbar g\ell _{x} \ell _{y} }{N} C_{0} X_{0}^{3} $ and $U_{2}=\frac{3\hbar g\ell _{x}^{2} \ell _{y}^{2}}{4N^{2}} C_{0} X_{0}^{5}$, respectively. In Eq. (\ref{HamMain}) we have also introduced a trapping potential $V_{tr}^{(pol)} (\mathbf{r})$ for the polaritons, which is assumed to  be parabolic, i.e., 
\begin{equation}
\label{Potents}
V_{tr}^{(pol)} (\mathbf{r})=\frac{m_{x} \omega _{x}^{2} x^{2} }{2} +\frac{m_{y} \omega _{y}^{2} y^{2} }{2}.
\end{equation}

Next, we use a mean-field approach to replace the corresponding polariton field operator $\hat{\Psi}\left(\mathbf{r}\right)$ by its average value $\left\langle \hat{\Psi }\left(\mathbf{r}\right)\right\rangle \equiv \Psi \left(\mathbf{r}\right)$, which characterizes the LB polariton wave function associated with the PolC structure.
By using Eq. (\ref{HamMain}), we obtain a governed equation for $\Psi \left(\mathbf{r}\right)$,
\begin{eqnarray}
\label{GPE1}
i\hbar \frac{\partial \Psi \left(\mathbf{r},t\right)}{\partial t}&=& \left\{-\frac{\hbar ^{2} \partial ^{2}}{2m_{x} \partial x^{2}}-\frac{\hbar ^{2} \partial ^{2}}{2m_{y} \partial y^{2}}+V_{tr}^{(pol)} \left(\mathbf{r}\right)\right.\phantom{00000} \\\nonumber
&+&\left. U_{1} \left|\Psi \left(\mathbf{r},t\right)\right|^{2} +U_{2} \left|\Psi \left(\mathbf{r},t\right)\right|^{4} 
\vphantom{\frac{\hbar ^{2} \partial ^{2}}{2m_{y} \partial y^{2}}} - i\gamma
\right\}\Psi \left(\mathbf{r},t\right),
\end{eqnarray}
which is a modified complex nonlinear Schr{\"o}dinger equation with a particle tensor mass, trapping potential, and cubic-quintic nonlinearities. In addition, we have introduced the last term in brackets that is responsible for nonequilibrium properties of polaritons. In particular, parameter~$\gamma = X_{0}^{2}\Gamma + C_{0}^{2}\gamma_{ph}$ characterizes the weak decay rate of the polariton number depending on photon leakage (parameter $\gamma_{ph}$) as well as on atomic decoherence (spontaneous emission rate $\Gamma$) for a coupled atom-light system in Fig.~\ref{PicLat}~(cf.~\cite{AlodjantsArakelianLeksinLaserPhysics}).

Various aspects of such an equation (in the isotropic case for $m_{x}=m_{y} \equiv m$) have been studied previously with respect to the analysis of SF behavior of atomic~\cite{AbdullaevGammalTomio} or photonic quantum ``liquids'' \cite{QuirogaTeixeiroMichinel}.

\begin{figure}
\includegraphics[width=8.4cm]{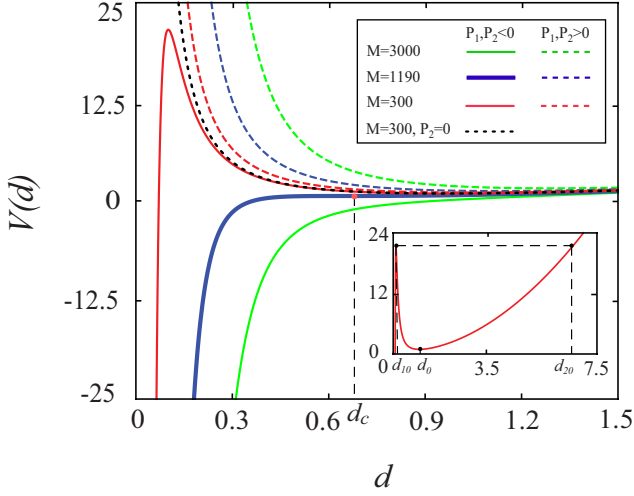}%for PDFLaTeX
\caption{\label{PicPot} (Color online) Effective potential $V(d)$ as a function of normalized width of the polariton wave function for positive and negative scattering lengths is shown by dashed and solid lines, respectively. The black dotted line characterizes a negative scattering length without quintic nonlinearity, $P_2 = 0$. The parameters used are $\Delta =0$, $N=50$, $n_{pol} ={N_{0}/M} =10$, $\ell_{x} =\ell_{y} =3\;\mu\text{m}$, $r_{0} =20\;\mu\text{m}$, and $g/2\pi =8.6\text{ GHz}$. In the insert $V\left( d \right)$ is plotted within a wide range of $d$; $M=300$. }
\end{figure}

In general it is useful to recast Eq.~(\ref{GPE1}) in terms of new coordinates  $\bar{x}=\sqrt{\frac{\mathstrut m_{x}}{m}}\:x$  and $\bar{y}=\sqrt{\frac{\mathstrut m_{y}}{m}}\:y$, introducing a new variable $\Psi \left(\mathbf{r},t\right)=\psi(\bar{x},\bar{y},t)e^{-\gamma t}$ that obeys the equation  
\begin{eqnarray}
\label{GPE2}
i\hbar \frac{\partial \psi}{\partial t}&=&\left\{-\frac{\hbar ^{2}}{2m} \left(\frac{\partial ^{2}}{\partial \bar{x}^{2}}+\frac{\partial ^{2}}{\partial \bar{y}^{2}} \right)+V_{tr}^{(pol)} \left(\bar{x},\bar{y} \right)\right. \nonumber\\
&+& \left. \bar{U}_{1} \left|\psi \right|^{2}+\bar{U}_{2} \left|\psi \right|^{4} \right\}\psi ,
\end{eqnarray}
where we have defined $\bar{U}_{1}=U_{1} e^{-2\gamma t}$ and  $\bar{U}_{2}=U_{2} e^{-4\gamma t}$; $m$~is an effective polariton mass. In Eq.~(\ref{GPE2}) the trapping potential given by Eq.~(\ref{Potents}) is represented  as ${V_{tr}^{(pol)}\left(\bar{x},\bar{y} \right)=\frac{m}{2} \left(\omega _{x}^{2} \bar{x}^{2}+\omega _{y}^{2} \bar{y}^{2} \right)}$. Thus, in the presence of the polariton number decaying we transform CNLSE to ``usual'' NLSE for wave function $\psi (\bar{x},\bar{y},t)$ with time-dependent nonlinear parameters $\bar{U}_{1}\left( t \right)$ and  $\bar{U}_{2}\left( t \right)$~(cf.~\cite{AndersonLisakReichelPRA,KohlerSols}). It is worth noticing that $U_{1}(t)$ and $U_{2}(t)$ are diminishing in time with different rates.  

Below we study ground-state properties of PolC by means of the variational approach for the solution of Eq.~\eqref{GPE2}. In  particular, we take the Gaussian trial function 
\begin{eqnarray}
\label{WaveFunc}
\psi (\bar{x},\bar{y},t)&=&\sqrt{\frac{N_{0} }{\pi R_{x} R_{y} } }\\
&\times & \exp \left[-\frac{\bar{x}^{2} }{2R_{x}^{2} } -\frac{\bar{y}^{2} }{2R_{y}^{2} } +\frac{i\bar{x}^{2} b_{x} }{2} +\frac{i\bar{y}^{2} b_{y} }{2} \right] \nonumber
\end{eqnarray}
for describing the quantum mechanical macroscopic ground state of LB polaritons. In the \textit{zero temperature} limit the wave function $\Psi \left(\mathbf{r}\right)$ is relevant to the description of LB polariton BEC that can occur in the limiting case at full thermal equilibrium; $N_{0} $ is the average total number of polariton particles. In this case one can assume that $N_{0} = n_{pol} \cdot M$.

The time-dependent function $R_{x,y}(t)$ determines the width of a wave function, and $b_{x,y}(t)$ characterizes a related wave function curvature. 
For further processing it is useful to introduce new dimensionless variables for the wave function width $d_{x,y}={R_{x,y} \mathord{\left/ {\vphantom {R_{x,y} r_{0}}} \right.} r_{0}}$, the rescaled time $\tau$~$=$~$\omega_{0}t$, and the decay rate $\Upsilon=\gamma/\omega_{0}$, with the characteristic length scale $r_{0}$~$\equiv$~$\sqrt{{\hbar \mathord{\left/ {\vphantom{\hbar m\omega _{0}}} \right. } m\omega _{0}}}$ and the geometric mean of the harmonic oscillator (trapping) frequency $\omega_{0}=\sqrt{\mathstrut \omega_{x} \omega_{y}}$.

For the 2D configuration of PolC illustrated in Fig.~\ref{PicLat}, we may take $r_{0} = 20\;\mu \text{m}$ and $\omega _{0} /2\pi \approx 7.5\text{ GHz}$ by referring to possible experimental parameters~\cite{BaliliHartwellSnoke}. 
With the corresponding Lagrangian for Eq. (\ref{GPE1}), 
\begin{eqnarray}
\label{Lagr}
L=&&\frac{\hbar N_{0}}{4} \sum _{\eta =x,y}\left\{ \vphantom{\frac{m\omega _{0}^{2} \lambda _{\eta}^{2} R_{\eta}^{2}}{\hbar}} R_{\eta }^{2} \frac{db_{\eta}}{dt}+\frac{\hbar}{m} \left(\frac{1}{R_{\eta }^{2} } +R_{\eta}^{2} b_{\eta}^{2} \right) \right. \nonumber\\
&+&\left. \frac{m\omega _{0}^{2} \lambda _{\eta}^{2} R_{\eta}^{2}}{\hbar} \right\}+\frac{N_{0}^{2} \bar{U}_{1}}{4\pi R_{x} R_{y}}+\frac{N_{0}^{3} \bar{U}_{2}}{9\pi ^{2} R_{x}^{2} R_{y}^{2}},
\end{eqnarray}
we can derive a set of coupled nonlinear equations for the wave function widths in $\bar{x}$ and $\bar{y}$ coordinates,
\begin{subequations}
\label{OscNorm}
\begin{eqnarray}
\label{OscNorm1}
\ddot{d}_{x} &=&\frac{1}{d_{x}^{3}}-\lambda _{x}^{2} d_{x}+\frac{P_{1}}{d_{x}^{2} d_{y}}+\frac{P_{2}}{d_{x}^{3} d_{y}^{2}},\\
\label{OscNorm2}
\ddot{d}_{y} &=& \frac{1}{d_{y}^{3}}-\lambda _{y}^{2} d_{y}+\frac{P_{1}}{d_{y}^{2} d_{x}}+\frac{P_{2}}{d_{y}^{3} d_{x}^{2}},
\end{eqnarray}
\end{subequations}
with $\lambda_{x,y}=\sqrt{\mathstrut \omega_{x,y} /\omega_{y,x}}$.

\begin{figure}
\includegraphics[width=8.4cm]{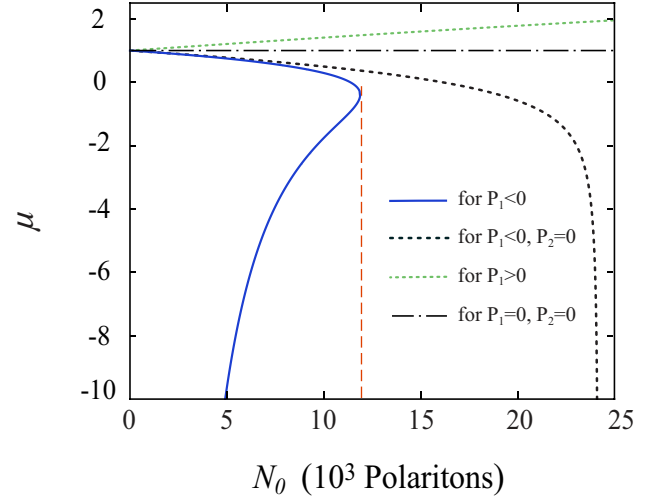}%for PDFLaTeX
\caption{\label{PicHimPot} (Color online) Dimensionless chemical potential $\mu$  versus the number of polaritons $N_{0}$. The vertical (red) dashed line corresponds to a critical number $N_{0 c} =11890$. The other parameters are the same as those in Fig.~\ref{PicPot}. The horizontal dashed-dotted line ($\mu = 1$) characterizes a related chemical potential for the ideal gas of noninteracting polaritons. Each curve is plotted with $d_{0}$ being in the steady state.\\}
\end{figure}

Two rescaled interaction parameters ${P_{1} \equiv N_{0} \bar{U}_{1} m / 2\pi \hbar ^{2}}$ and ${P_{2} \equiv 4N_{0}^{2} \bar{U}_{2} m / 9\pi^{2} \hbar^{2} r_{0}^{2}}$ are introduced for polariton-polariton two- and three-body scattering processes, respectively. Notice that, practically, the values of interaction parameters for PolC structures satisfy the inequality $\left|P_{2} \right| \ll \left|P_{1} \right|\le 1$ for a small number of atoms. On the other hand, $\left|P_{1} \right| \gg 1$ in the Thomas-Fermi limit, which implies a large number of microcavities such as $M \gg 10^{3}$. 

We examine equilibrium properties of polaritons which can be obtained on short time scales or for negligibly small rates of $\gamma $. Loosely speaking we are taking parameters $P_{1,2} $ in Eq.~\eqref{OscNorm} as a constant in time.

We have also assumed that, initially, quasiparticles are placed at rest, i.e., $\dot{d}_{x} (0)=\dot{d}_{y}(0)=0$. The equilibrium points $d_{x}=d_{x0}$ and $d_{y}=d_{y0}$ for wave function widths in two dimensions are determined in steady-state conditions:
\begin{equation}
\label{EquilibR}
\lambda _{x,y}^{2} d_{x,y0}=\frac{1}{d_{x,y0}^{3}} +\frac{P_{1} }{d_{x,y0}^{2} d_{y,x0} } +\frac{P_{2} }{d_{x,y0}^{3} d_{y,x0}^{2}},
\end{equation} 
which  can not be solved analytically in a general case. We first analyze the stability of PolC with some specific physically important limits. 

\subsection{\label{analysisofradialsymmetric} Symmetric (isotropic) PolC}
For a complete isotropic configuration of PolC, we can assume trapping potential frequencies to be equal, i.e., $\lambda _{x}=\lambda _{y}=1$, and take $d_{x,y} = d$. From Eq. (\ref{OscNorm}), we obtain a Newton-like differential equation, 
\begin{equation}
\label{SymmOsc}
\ddot{d}=\frac{1+P_{1}}{d^{3}}-d+\frac{P_{2}}{d^{5}},
\end{equation}
with an effective potential,
\begin{equation}
\label{SymmPot}
V(d)=\frac{1+P_{1}}{2d^{2}}+\frac{P_{2}}{4d^{4}}+\frac{d^{2}}{2},
\end{equation}
and a corresponding dimensionless chemical potential $\mu$,
\begin{equation}
\label{SymmHimPot}
\mu =\frac{1}{2} \left(\frac{1}{d^{2}} +d^{2} \right)+\frac{P_{1}}{d^{2}} +\frac{3P_{2} }{4d^{4} }.
\end{equation}

\begin{figure}
\includegraphics[width=8.4cm]{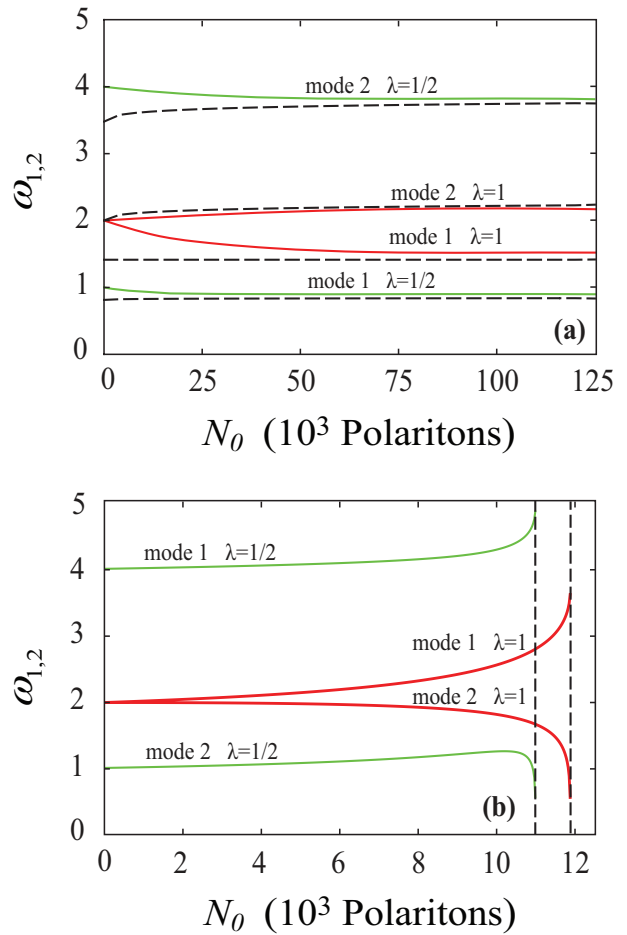}%for PDFLaTeX
\caption{\label{PicOsc} (Color online) Normalized low amplitude oscillation frequencies $\omega_{1,2}$ (in trapping frequency $\omega_{0}$ units) as a function of polariton particle number $N_{0}$ for (a) $P_{1} > 0$ and (b) $P_{1}<0$. The horizontal dashed curves in (a) characterize the Thomas-Fermi limit, while the vertical dashed lines in (b) correspond to a critical number of particles for which the collapse of wave function happens. Other parameters are the same as those in Fig.~\ref{PicPot}.}
\end{figure} 

In Fig.~\ref{PicPot} the dependencies for potential $V(d)$ as a function of the width $d$ of the PolC wave function are shown. A corresponding chemical potential behavior is represented in Fig.~\ref{PicHimPot}. 
For a positive polariton scattering length ($P_{1}>0$), an effective potential $V(d)$ has only one minimum $d_{0}$ and a supported polariton wave function is always stable. The quintic nonlinearity $P_2$ in this case does not play an important role. From Eq.~(\ref{SymmOsc}) we get an equilibrium value of wave function width $d_{0} \simeq \left(1+P_{1} \right)^{1/4}$ that corresponds to the chemical potential $\mu=\left(1+1.5P_{1} \right)/\left(1+P_{1} \right)^{1/2}$, as shown by the green (upper) dotted curve in Fig.~\ref{PicHimPot}. 

The situation dramatically changes if we deal with polaritons with a negative scattering length, $P_{1} < 0$.
In this case the atom-field coupling constant $g$ should be negative too.
A polariton wave function is found to be stable if we completely neglect three-body polariton interactions, i.e., $P_2 = 0$, as shown by the black dotted curves of Figs.~\ref{PicPot} and~\ref{PicHimPot}, respectively. 
Roughly speaking, for $P_{2} = 0$, the effective potential $V(d)$ has one equilibrium point defined as $d_{0} \simeq \left(1-\left|P_{1} \right|\right)^{1/4}$.

In the presence of quintic nonlinearity, PolC becomes unstable and the corresponding wave function collapses, as shown in the solid curves of Fig.~\ref{PicPot} and Fig.~\ref{PicHimPot}.
In particular, a critical polariton number is found as $N_{0 c} = 11890$, which corresponds to nonlinear parameters $\left|P_{1 c} \right|=0.4915$ and $\left|P_{2 c} \right|=0.1396$; all this relates to the blue solid (bold) curve~in~Fig.~\ref{PicPot}.
The corresponding critical width of a polariton wave function is $d_{0c} =0.6417$. 
It is interesting to note that such a behavior of the polariton wave function  is commonly inherent to BECs in higher dimensions ~\cite{AbdullaevGammalTomio}. 

For a number of cavities, such as $M=M_{c}$, there exists one metastable point~$d_{c}$ for a polariton wave function that characterizes the bending of the effective potential~$V(d)$, as shown by the blue solid (bold) curve of Fig.~\ref{PicPot}.
As the number of cavities increases, $M>M_{c}$, our PolC is unstable and the related wave function $\Psi \left( \mathbf{r} \right) $ collapses, as shown by the green solid (lower) curve of Fig.~\ref{PicPot}. 
For a smaller number of microcavities, $M<M_{c}$, there exist two equilibrium points $d_{10}$ and $d_{0}$ for a polartion ``cloud'' behavior, as shown in the inset of Fig.~\ref{PicPot}.
One of these two equilibrium points , $d_{10}$, is unstable.
A LB polariton wave function behaves unstable and tends to collapse if we go to the left from this point, i.e., for $d\le d_{10}$.
 On the other hand, a polariton cloud oscillates within the region $d_{10} \le d\le d_{20}$. 

\subsection{\label{analysisofpolaritonbec} Anisotropic PolC: small amplitude oscillations}

Expanding Eqs. (\ref{OscNorm}) around the equilibrium points $d_{x0}$ and $d_{y0}$ one can easily find low-energy-excitation frequencies for LB polaritons in a PolC structure as
\begin{eqnarray}
\label{Freq}
\omega _{1,2} &=&\sqrt{2} \omega _{0} \left[
\vphantom{\sqrt{\left(P_{2}^{33} \right)^{2}}}
\left(\lambda _{x}^{2} +\lambda _{y}^{2} -P_{1}^{13} -P_{1}^{31} \right)\right.\\ \nonumber
& \pm &\left.\sqrt{\left(\lambda _{x}^{2} -\lambda _{y}^{2} +P_{1}^{13} -P_{1}^{31} \right)^{2} +4\left(P_{1}^{22} +P_{2}^{33} \right)^{2}} \right]^{1/2} ,
\end{eqnarray}
where $P_{1}^{ij} =\frac{P_{1}}{4d_{x0}^{i} d_{y0}^{j} } $ and $P_{2}^{ij} =\frac{P_{2}}{2d_{x0}^{i} d_{y0}^{j} }$.
Two types of orthogonal oscillation modes are determined by two signs in Eq. (\ref{Freq}). 
In Fig.~\ref{PicOsc}(a) we plot the dependencies of small oscillation frequencies $\omega_{1,2}$ (in~$\omega_{0}$~units) as a function of polariton particle number $N_{0}$ for the case $P_1 > 0$. The horizontal dashed curves characterize the Thomas-Fermi limit. In particular, oscillation frequencies $\omega_{1,2}$ approach $\omega_{1} \simeq \omega _{0} \sqrt{2}$ and $\omega_{2} \simeq \omega_{0} \sqrt{2} \left(2+\frac{P_{2}}{d_{0}^{6}} \right) ^{1/2}$ for a symmetric case when wave function spatial widths $d_{x,y}$ and trapping frequencies are the same, i.e. $d_{x,y0} =d_{0}$ and $\lambda_{x} =\lambda_{y} =1$, respectively. 
Instead, for a negative scattering length, i.e., for $P_{1} < 0$,  the characteristic oscillations are limited by critical value $N_{0c}$ for which collapse of PolC wave function occurs~(cf.~\cite{PerezGarciaMichinel}). 

\begin{figure}
\includegraphics[width=8.4cm]{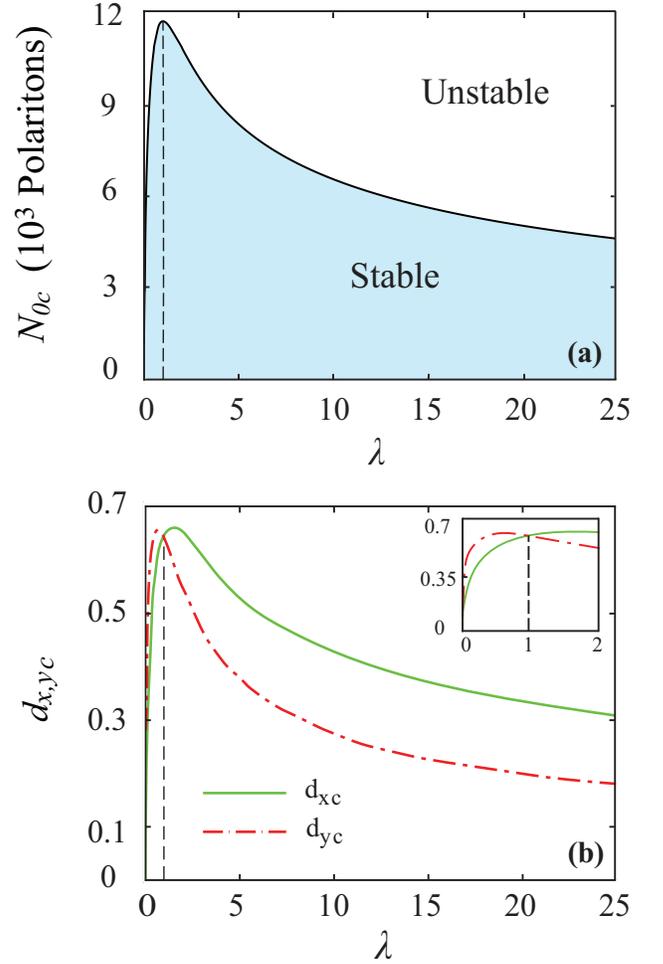}
\caption{\label{PicNiWotLambda} (Color online) (a) Critical number $N_{0c}$ of polaritons at the ground state of PolC and (b) the corresponding wave function widths $d_{x,yc}$ against a normalized trapping frequency parameter $\lambda$ for a negative scattering length ($P_{1} <0$). The vertical dashed line corresponds to the symmetric case with $\lambda =1$. Other parameters are the same as those in Fig.~\ref{PicPot}. The shaded region characterizes a stable domain for PolC structure wave function.}
\end{figure} 

For a nonsymmetric case we choose the parameters $\lambda \equiv \lambda _{y}={1\mathord{\left/ {\vphantom {1 \lambda _{x}}} \right.} \lambda _{x}}$ and demonstrate in Fig.~\ref{PicNiWotLambda} the dependence of the critical number of polaritons $N_{0c}$ and related critical widths $d_{x,yc}$ of a wave function on parameter $\lambda$ for negative scattering length. In both cases the particle number $N_{0 c}$ as well as wave function widths $d_{x,yc}$ diminish due to symmetry properties of Eq. (\ref{EquilibR}). From Fig.~\ref{PicNiWotLambda}(a) it is clearly seen that the maximal value $N_{0 c}=11890$ is obtained for a radially symmetric polariton cloud with trapping frequencies $\omega_{x} =\omega_{y}$ ($\lambda =1$). On the other hand, two limits $\lambda<<1$ $\left( \omega_{y}<<\omega_{x}\right) $ and $\lambda >>1$ $(\omega_{y}>>\omega_{x})$ correspond to highly anisotropic traps which physically correspond to the reduction of a 2D lattice to a 1D spatially periodic structure~(see Fig.~\ref{PicLat}). The collapse of a wave function happens for parameters of the system belonging to the domain at the outside of the shaded region in Fig.~\ref{PicNiWotLambda}(a). It is interesting to note that the extreme width of the wave function in one spatial dimension is achieved for the nonsymmetric case for which $\omega_{x} \ne \omega_{y}$. In particular, $d_{xc}$ reaches its maximal value $d_{xc,\, \max} =0.6618$ at $\lambda =1.518$. The same magnitude of $d_{yc,\, \max} $ can be obtained in another limit of polariton trapping frequencies for $\lambda =0.659$.

\subsection{\label{NLDyn} Dissipative dynamics}

\begin{figure}
\includegraphics[width=7cm]{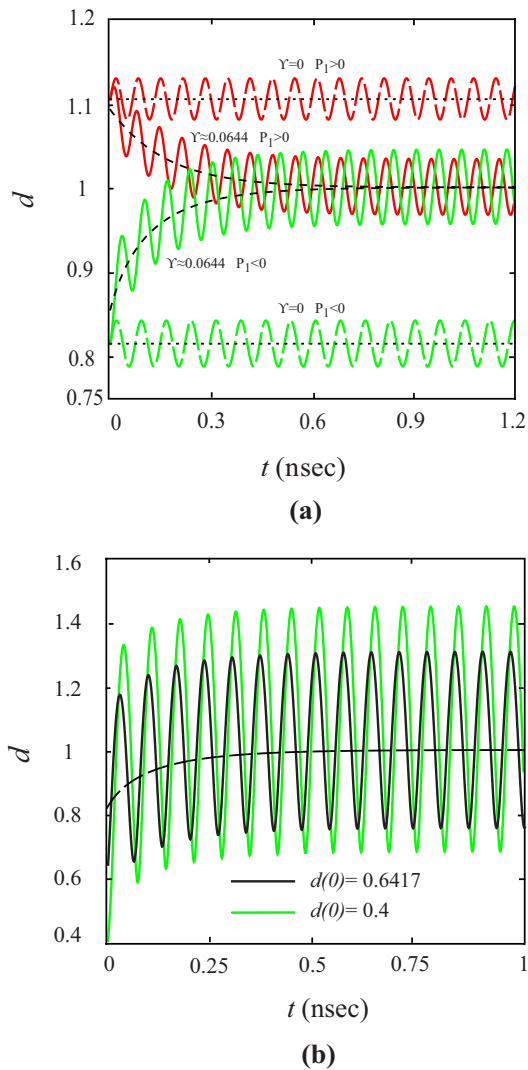}
\caption{\label{PicOscPozINeg} (Color online) Wave function width $d$ against time $t$. The parameters are the following: (a) $N_{0}$~$=$~$10000$ ($\left|P_{1}\right|$~$=$~$0.414$, $\left|P_{2}\right|$~$=$~$0.099$),  $\dot{d}(0)$~$=$~$0.05$, $d_{0}$~$=$~${ 1.106}$ for positive scattering length and $d_{0} =0.813$ for negative scattering length;  the dotted black lines  correspond to steady state solutions; (b) ${N_{0}=N_{0 c}=11890}$ ($P_{1}$~$=$~$P_{1 c}$~$=$~$-0.4915$, $P_{2}$~$=$~$P_{2 c}$~$=$~$-0.1396$), $\dot{d}(0)$~$=$~$0.5$,~${\Upsilon \approx 0.0644}$ $\left( \gamma/ 2\pi \approx 0.4805 \text{ GHz}\right)$.}
\end{figure} 

Let us examine nonequilibrium properties of a polariton system. Notably, since the relation $\gamma <<g, \omega_{1,2}$ is fulfilled, the adiabatic approximation is valid for the problem under discussion (cf.~\cite{KohlerSols}). A particle number decaying that is inevitable in the general case due to polariton interaction with the environment enables us to change the physics of PolC dynamics sufficiently. Figure~\ref{PicOscPozINeg} demonstrates typical temporal dynamics of  PolC wave function width $d=d_{x} = d_{y}$ in the adiabatic limit for the symmetric case $\lambda=1$. The initial conditions taken for Fig.~\ref{PicOscPozINeg}(a) are very close to steady-state point $d_{0}$ described by Eqs.~(\ref{EquilibR}) --- black dotted lines. Completely neglecting the decay rate, the polaritonic system exhibits small amplitude oscillations for positive (the upper dashed curve) and  for negative (the lower dashed curve) scattering lengths. In the presence of polariton decaying the steady-state levels (dotted lines~in~Fig.~\ref{PicOscPozINeg}) of width~$d$ are adiabatically  shifted to the value $d_{st} \simeq 1$ which occurs due to the diminishing of governed parameters~$P_{1,2}$. A newly settled regime of small-amplitude oscillations around $d_{st}$ can be described by Eq.~(\ref{Freq}) as well. By using a mechanical analogy (see, e.g., \cite{Arnold}) we focus on the fact that since the action is an adiabatic invariant, the energy of small-amplitude oscillations remains proportional to the frequency of oscillations by slow decaying PolC parameters.   

The influence of a decay rate on polariton dynamics becomes more evident if we consider a PolC with negative scattering length initially prepared in an unstable (collapsing) region. In Fig.~\ref{PicOscPozINeg}(b) the nonequilibrium dynamics for PolC wave function width is shown. The initial conditions are taken for critical width  ${d(0)=d_{c} \simeq 0.6417}$, which is shown by the blue solid (bold) curve in Fig.~\ref{PicPot} and for ${d(0)=0.4<d_{c}}$. A collapse of wave function occurs if we neglect decaying of PolC particles. However, for a small but finite  decay rate $\gamma$ the system undergoes a transition from an unstable region to a stable domain of small amplitude oscillations, avoiding a collapse of wave function. One can also give another explanation for this. Actually, since parameters $P_{1} \propto e^{-2\gamma t}$ and $P_{2} \propto e^{-4\gamma t}$ vary in time with different rates, a contribution of quintic nonlinearity in the temporal dynamics of the system vanishes much faster. Hence, the domain of PolC wave function collapse should vanish as well. For a much smaller value of the initial wave function width  $d$ the collapse can be escaped for much larger values of  decay rate $\gamma$. However, for large decay rates such as $\gamma \gtrsim g,\omega _{{ 1,2}}$ that are beyond the adiabatic approximation, our approach based on a variational method becomes inadequate.

\section{\label{conclus}CONCLUSIONS}

In the paper, we consider a 2D spatially periodic structure, coined as a polaritonic crystal, for observing macroscopic properties for coupled atom-field states (polaritons) in the lattice at the zero temperature limit.
 Under the tight-binding approximation such a system realizes weakly coupled cavities containing a small amount of two-level atoms which interact with quantized cavity modes. We have shown that the dynamics of the polaritons in the lattice is much richer if it is beyond a typically used low-density limit. 
First of all, we have studied two- and three-body polariton-polariton scattering parameters by means of the Holstein-Primakoff approach.
We have shown that two-body polariton scattering dominates in the positive atom-field detuning domain that corresponds to atomlike LB polaritons.
As a sequence, we consider macroscopic properties of such polaritons as a whole at the bottom of the dispersion curve in the continuous limit of a spatially periodic (lattice) structure.
A variational approach  is used to study the related  widths, chemical potential, and characteristic frequencies of  PolC ground-state wave functions around the equilibrium points. In particular, we consider the polariton decay rate  $\gamma $, which is essentially smaller than other characteristic parameters, such as atom-field coupling strength $g$ and effective trapping frequency $\omega _{0} $. Physically it means that the crucial parameters $P_{1,2} $ that characterize polaritonic nonlinearity vary adiabatically slowly in time. This approach is justified by supporting sufficiently low temperatures of a polaritonic system and by exploring cavities with a high \textit{Q}-factor for PolC design purposes. Our results reveal the fact that an unstable ground state in two dimensions is supported beyond the critical number of polaritons and low-excitation-density limit for negative scattering length. Simulation of small-amplitude nonequilibrium  (dissipative) dynamics in the presence of condensate particle decaying reveals new features in PolC  behavior. For some values of initial conditions belonging to the unstable domain, the polaritonic system adiabatically crosses a collapsing region and reaches a stable regime of small-amplitude oscillations. Thus, for a negative scattering length the decay process in some cases prevents PolC wave function  from collapse and fragmentation.

\begin{acknowledgments}
This work was supported by RFBR Grants No. 10-02-13300 and No. 11-02-97513 and by the Russian Ministry of Education and Science under Contracts No. $\Pi569$, No. $\Pi335$, and No. 14.740.11.0700.
\end{acknowledgments}

\end{document}